\documentclass[10pt]{article}

\usepackage{caption}
\usepackage{graphicx}
\usepackage[margin=1in]{geometry}   
\usepackage{placeins}
\usepackage{float}
\usepackage[utf8]{inputenc}
\usepackage[T1]{fontenc}

\begin{document}

\title{\vspace{-10pt} Encouraging Equitable Bikeshare: \\
       Implications of Docked and Dockless Models for Spatial Equity}
       
\author{\begin{tabular}{cc}Simon Couch\hspace{30pt} & \hspace{30pt} Heather Kitada Smalley \\ \textit{\small couchs@reed.edu}\hspace{30pt}  & \hspace{30pt}\textit{\small dr.kitadasmalley@gmail.com} \\ \small Reed College \hspace{30pt} & \small \hspace{30pt}Willamette University\footnote{This work was completed while Dr. Smalley was a Visiting Assistant Professor at Reed College.} \end{tabular}}

\date{}

\maketitle

\section*{Abstract}\label{sec:abstract}

The last decade has seen a rapid rise in the number of bikeshare programs, where bikes are made available throughout a community on an as-needed basis. Given that many of these programs are at least partially publicly funded, a central concern of operators and investors is whether these systems operate equitably. Though spatial equity has been well-studied under the docked model, where bikes are picked up and dropped off at prespecified docking stations, there has been little work examining that of the increasingly popular dockless model, where bikes can be picked up and dropped off from anywhere within an operating area. We explore comparative equity in spatial access to bikeshare services under these two models by collecting spatial data on 45,935 bikes from 73 bikeshare systems using a novel querying approach (with generalizable and freely available source code), and joining this data with newly-available sociodemographic data at the census tract level. Using Poisson count regression, we perform the first comparative analysis of the two docking approaches, finding that dockless systems operate more equitably than docked systems by education, but do not differ in spatial access by socioeconomic class.

\clearpage

\section{Introduction}\label{sec:data}

Several decades of research in urban planning has shown that opportunity for active transportation is deeply connected to the health of communities \cite{ellin2012, handy2005, katz1994}. Creating this opportunity is known to require active investment and planning \cite{litman2015, speck2013}, but these efforts to establish and maintain suitable infrastructure consistently have been shown to have tangible and substantial impacts on participation in active transport \cite{le2018, us2015, ricci2015}. Bikeshare, where bikes are made available on an as-needed basis throughout a community, holds significant promise as an element of this infrastructure. Bikeshare is believed to be a potential solution to the first and last mile problems of more well-established public transit networks, and even to serve as a standalone feature of public transit \cite{fishman2016, martin2014, pucher2011}. As a result, the popularity of, and investment in, bikeshare systems has exploded in the past two decades; in 2001, there were five bikeshare systems operating in five European countries, and as of the writing of this paper, there are over 1200 systems operating in 91 countries all over the world \cite{wiki:bikeshare, midgley2011}.

Given the history of other public transportation options underserving traditionally marginalized groups, though, there has been concern about risks of inequity in these bikesharing services. This potential for inequity has been a small but growing area of research, and indeed, higher-income, college-educated whites have been shown to be better served by these systems \cite{fishman2016, mcneil2018, mooney2019, ricci2015}. Generally, these studies have focused on docked bikeshare systems, where bikes are made available at a set of predetermined stations, and users can take a bike from one station to another. Studies on the spatial equity of these systems have uncovered modest but consistent disparities in access to station locations based on sociodemographic predictors \cite{hosford2018, meng2018, ursaki2015}, sparking a subfield of research on methodology to distribute these stations more equitably \cite{bhuyan2019, conrow2018, griffin2018}. Further, this docked strategy can be costly and inflexible---stations generally cost \$30,000 to \$50,000 a piece, and are generally regarded as permanent features of a bikeshare system \cite{shaheen2014}. 

In response to these issues, there has been a rapid rise in dockless bikeshare systems, where users can locate and unlock bikes using smartphone applications, and then ride the bike to any other location within the bikeshare system's service area. First rising to popularity in China, and then making their way to North America in 2017, dockless bikeshare systems are popping up across many North American cities \cite{mooney2019, sin2017}. There are arguments in both directions on the implications of the dockless model for the equity of bikeshare systems. On one end, inequities are no longer ``built-in'' to these systems, in the sense that the placement of station locations, which is known to concentrate in advantaged neighborhoods, is no longer consequential for accessibility. Further, due to the savings resulting from the lack of need to build docking stations, dockless systems have been reported to be launching with higher bike-to-resident ratios than docked systems \cite{kroman2017}, potentially allowing for access for a greater diversity of users. However, it is also plausible that dockless bikes will end up in neighborhoods housing wealthier and more educated residents, who currently use bike sharing services at a higher proportion \cite{mooney2019}. Further, it is not known the effect that the process of rebalancing, where bikes are redistributed according to projected demand, has on the spatial equity of these systems \cite{dechardon2016}. A recent study on dockless systems found that modest spatial inequities among socioeconomic lines existed in Seattle, but that disparities in access by gentrification risk or racial makeup were not significant \cite{mooney2019}.

In general, existing studies in equity of bikeshare systems have been 1) limited to one or a small number of cities, 2) are based on survey data, and 3) focus only on the docked bikeshare model. There are notable exceptions to each these generalities. In regard to sample diversity, in 2015, Smith et al. gathered data on 42 bikesharing systems making up 2,137 docking stations throughout North America. Also, McNeil et al. surveyed 56 bikeshare \textit{operators}, though surveys collected from bikeshare \textit{users} came from only three cities \cite{smith2015}. In a similar study, Leister et al. surveyed 23 bikeshare system operators about several dimensions of access and equity \cite{leister2018}.  Several papers reporting on one or a small number of cities acknowledged the potential for lack of generalizability to cities with largely different populations. There is a slightly larger presence of studies utilizing non-survey data: Smith et al. utilized coordinates of stations merged with sociodemographic data from the American Community Survey in 2015 \cite{smith2015}, Ursaki et al. implemented a similar approach on seven cities in 2015 \cite{ursaki2015}, Hosford et al. utilized similar tactics on five canadian cities in 2018 \cite{hosford2018}, and Mooney et al. analyzed data on the coordinates of undocked bikes. As for the dockless model, Mooney et al. conducted a study in 2019 on dockless bikeshare in Seattle, finding modest spatial inequities in access to bikeshare bikes between incomes and education levels, but not detecting significant disparities by gentrification risk or racial makeup \cite{mooney2019}. Since this work was focused on one city, the question remains of the \textit{comparative} inequities of the docked and dockless models. Aside from Mooney et al., to our knowledge there is no existing work on the equity of bikeshare systems under the dockless model---this leaves the question of which of docked or dockless bikeshare systems is more equitable in terms of spatial access.

Our work extends the current literature in several ways. For one, we collect data from 73 bikeshare systems across the United States, containing information on 45,935 docked bikes as well as bikes not docked at a station (``free bikes''), making up the most diverse existing bike-level sample data to our knowledge. This data contains information on bikes operating under both docked and dockless systems, allowing not only for further study on spatial equity of dockless systems but allowing for direct comparison of the docked and dockless approaches. Our data collection methods are also generalizable and can contribute to future research in this field---we developed a publicly available R software package to query live bikeshare feed data \cite{gbfs}, and our source code applying this tool is also freely available\footnote{Source code is available at: github.com/simonpcouch/bikeshare}.

\section{Methods}\label{sec:methods}

In November 2015, the North American Bikeshare Association announced the General Bikeshare Feed Specification (\textit{gbfs}), an open data standard for bikeshare owners and operators to release real-time data on bikeshare systems. Intended to provide data in an easily accessible format for transportation based-apps and other integrated softwares, this standard has since been adopted by over 200 bikeshare systems at the time of writing \cite{nabsa_repo, nabsa_gbfs}. The feeds contain live information on, among other things, locations of free bikes, coordinates of docking stations, and pricing plans.

While this standard makes real-time data publicly available through live \textit{.json} feeds, this specification does not provide historical information about bikeshare systems. For this reason, we developed an \textit{R} package, freely available in the \textit{Comprehensive R Archive Network}, to query and archive this feed data in a rectangular format \cite{gbfs}. This software package allows \textit{R} users to accumulate data on bikeshare systems over time, allowing for the potential to generate datasets from open-access data that are richer and more diverse than utilized in previous research on bikeshare systems. 

\begin{figure} [!htb]
    \centering
    \includegraphics[width=.7\linewidth]{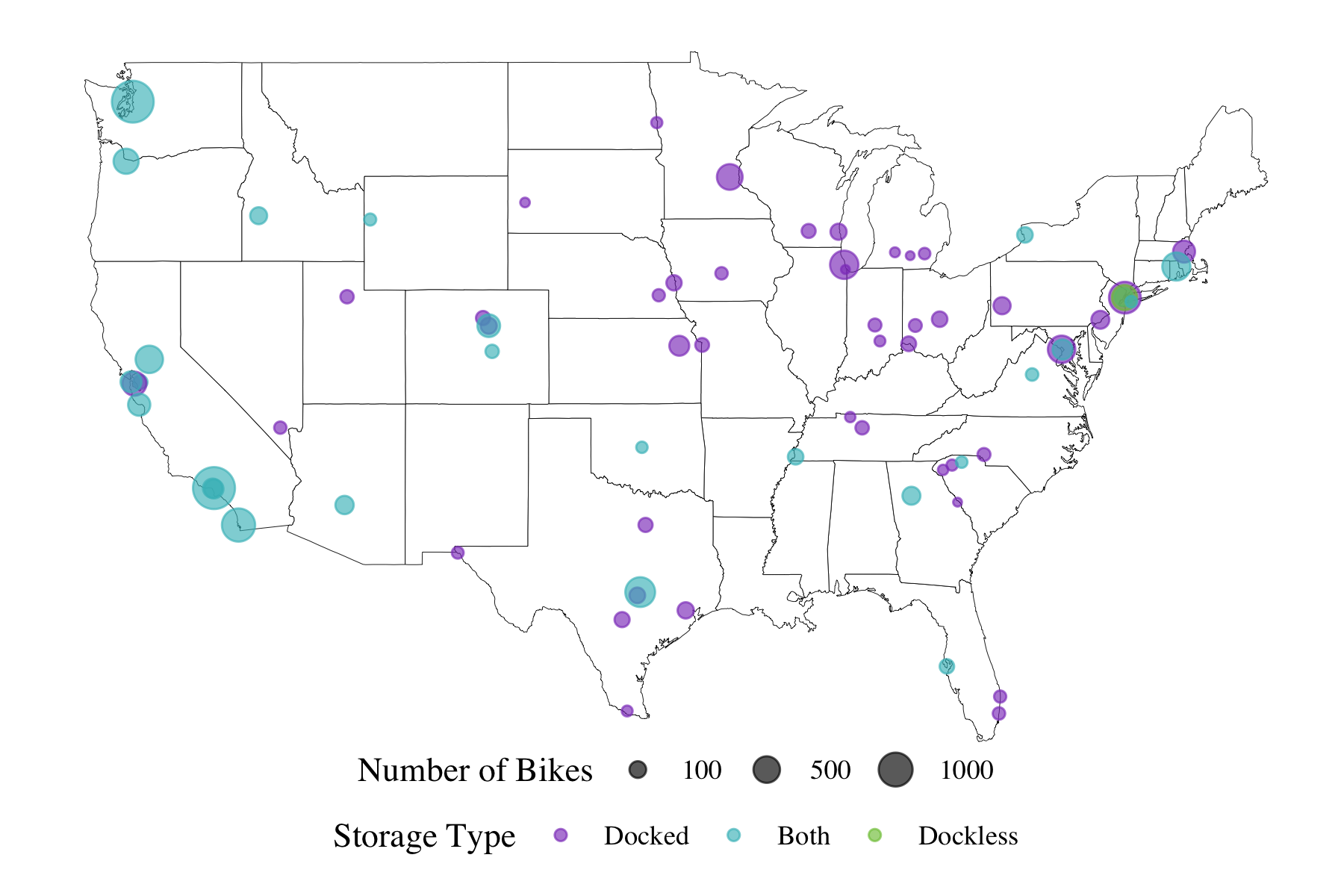}
    \caption{The geographic distribution of our sample data containing 45,935 unique stations and free bikes across 73 bikeshare systems operating in the United States.}
    \label{fig:us_bikes}
\end{figure}

We made use of this package to gather data from all bikeshare systems operating in the United States releasing \textit{gbfs} feeds at the time of writing. Specifically, we queried all feeds containing information on either the geographic coordinates of free bikes or docked bikes by iterating our querying script over all operating bikeshare systems in the U.S., standardizing dataset formatting (the \textit{gbfs} allows for slight deviances from the standard), and then binding the rows of each dataset together. Joining this data together, we gathered 45,935 observations from 73 unique bikeshare systems in the United States. The geographic distribution of this data is shown in Figure \ref{fig:us_bikes}.

% latex table generated in R 3.5.3 by xtable 1.8-4 package
% Thu May 16 09:59:03 2019
\begin{table}[ht]
\centering
\begin{tabular}{lll|lll}
\hline
 Storage Type & Bike Count & Number of Systems & \multicolumn{3}{c}{Percentile Bikes Per System} \\ 
  \cline{4-6}
 &  &  & 25\% \hspace{10pt} & 50\% \hspace{10pt} & 75\% \hspace{10pt} \\ 
   \hline
Dockless & 7108 & 28 & 5.5 & 35 & 311.5 \\ 
  Docked & 38827 & 72 & 51.5 & 156 & 364.5 \\ 
  \end{tabular}
  \caption{Summary statistics on the data resulting from our querying procedure. While there are significantly less dockless systems in our data, the number of bikes is comparable. Though there seem to be less dockless bikes per system, this difference is not statistically significant ($p = .119$).}\label{tab:sum_table}
\end{table}

To understand the relationship that the demographic characteristics of the neighborhoods these bikeshare systems inhabit have with the systems themselves, we joined the dataset described above with Opportunity Insights' Table 9: Neighborhood Characteristics by Census Tract. Opportunity Insights is a research institute of Harvard University that, among other things, provides detailed data about sociodemographic characteristics and economic mobility at the census tract level \cite{chetty2018, opp_insights}. To join this data, we needed to collapse the level of observation of our bikeshare data from each row representing a bike to each row representing a census tract. To do so, we reverse geocoded every observation in the dataset, extracting the census tract from the geographic coordinates of bikes by matching bike coordinates with the geographical boundary they belong to, and then summarizing the data such that each row was a unique census tract and count of free bikes or stationed bikes. Lastly, to address zero-inflation resulting from the large proportion of census tracts in the U.S. without bikeshare programs, we filtered out counties that contained counts of zero bikes for all census tracts they contained.

Finally, using this dataset, we fitted a Poisson regression model to predict the count of bikes per census tract based on the sociodemographic characteristics (scaled from 0 to 1 to allow for comparison of model coefficients) of the census tract. Estimating the log of the number of bikes in the census tract, Poisson models are appropriate for fitting counts, as opposed to any continuous value as in the more conventional least-squares regression setting. Though the coefficients themselves are only additive on the log scale, making them difficult to reason about, the exponentiated coefficients represent a multiplicative factor on the predicted response value. Then, to understand the relationship in spatial equity between docked and dockless systems, we fit an interaction term giving whether the predicted count was for the number of docked bikes or free bikes; a statistically significant coefficient in any of these interaction terms signifies that the docked and dockless models notably differ in some dimension of spatial equity.

The source code for the software, analysis scripts, and data described above is freely available\footnote{Source code is available at: github.com/simonpcouch/bikeshare}\cite{gbfs}.

\section{Results}\label{sec:results}

% latex table generated in R 3.5.3 by xtable 1.8-4 package
% Mon May 20 13:37:06 2019
\begin{table}[ht]
\centering
\begin{tabular}{rlrrll}
  & Predictor & Coefficient & exp(Coefficient) & p-Value & Significance \\ 
  \hline
1 & Intercept & -3.209 & 0.040 & $<$ .001 & *** \\ 
  2 & \% College-Educated & 3.799 & 44.654 & $<$ .001 & *** \\ 
  3 & \% in Poverty & 3.815 & 45.368 & $<$ .001 & *** \\ 
  4 & \% Nonwhite & -0.103 & 0.902 & 0.060 &  \\ 
  5 & Population Density & -1.326 & 0.265 & $<$ .001 & *** \\ 
  6 & Job Density & 0.730 & 2.075 & 0.138 &  \\ 
  7 & Docking Type & 0.922 & 2.515 & $<$ .001 & *** \\ 
   \hline
8 & \% College-Educated  * Docking Type & 0.557 & 1.746 & $<$ .001 & *** \\ 
  9 & \% in Poverty  * Docking Type & -0.079 & 0.924 & 0.234 &  \\ 
  10 & \% Nonwhite  * Docking Type & 0.612 & 1.844 & $<$ .001 & *** \\ 
  11 & Population Density  * Docking Type & 6.317 & 553.983 & $<$ .001 & *** \\ 
  12 & Job Density  * Docking Type & 2.444 & 11.519 & $<$ .001 & *** \\ 
   \hline
\end{tabular}
\caption{A Poisson regression model predicting the number of bikes per census by scaled sociodemographic characteristics of the neighborhood, presence of bikeshare in the county, and type of bikeshare service. Note that the coefficients are additive to the log predicted count, and the exponentiated coefficients are multiplicative to the actual count. Significance levels * = .1, ** = .05, and *** = .01}\label{tab:model}
\end{table}

As reflected in previous literature, several sociodemographic characteristics are intertwined with spatial access to bikeshare. In this sample, more college-educated (p < .001) census tracts tend to have greater spatial access to bikeshare services, reaffirming previous findings in the existing literature \cite{fishman2016, mcneil2018, ricci2015}. Though the model indicates slights disparities in spatial access to bikeshare by race, this effect is not statistically significant (p = .06), mirroring existing work on dockless systems \cite{mooney2019}. However, when controlling for these factors, as well as population density and job density, we find that spatial access is significantly greater in poorer communities (p < .001). This finding is reflected in some recent work \cite{wang2018}, yet refuted in most. Differences in modeling and significance testing approaches are likely relevant here---due to how deeply related race, class, and education are in the United States, fitting to aggregrate measures rather than specific demographic characteristics, as well as testing for effects of single measures without controlling for others, consistently shows that whiter, wealthier, and more educated communities are better served by bikeshare programs \cite{babagoli2019, mcneil2018, mooney2019, ricci2015, smith2015, ursaki2015}. In this way, our findings are not necessarily contradictory, in that we find that contextualized measures of demographic characteristics of American communities offer a different perspective than these measures in isolation.

The \textit{Docking Type} term, as well as its interaction terms in rows $8-12$ of the table, represents the change in predicted count if the model is fitting the count of docking stations rather than that of free bikes. That is, an exponentiated coefficient above $1$ signifies that the influence of the relevant predictor is more positive when the model is predicting the count of docked bikes rather than dockless bikes. This portion of the model allows us to examine the comparative spatial equity of programs operating under the docked and dockless models. Our model shows that, on one end, the predicted number of bikes in majority non-white communities is larger for docked systems than dockless systems (p < .001). (Note that we have controlled for the greater number of bikes in docked systems in general.) That is, the docked model seems to perform better for spatial equity in regard to race than the dockless model. In regard to educational boundaries, the converse seems to be true---the estimation for the number of bikes in more educated communities is greater under the docked model than the dockless model, suggesting that the dockless model provides better access for less-educated neighborhoods than the docked model (p < .001). The interaction term for the comparative impacts of these two models with regard to poverty is not statistically significant (p = .234), supporting the assertion that the models do not differ in spatial access with regard to socioeconomic class.

% ----------------------------------------------------------------------
% When the model table is reran, drop this caption in between
% tabular and table

%\caption{A Poisson regression model predicting the number of bikes per census by scaled sociodemographic characteristics of the neighborhood, presence of bikeshare in the county, and type of bikeshare service. Note that the coefficients are additive to the log predicted count, and the exponentiated coefficients are multiplicative to the actual count. Significance levels * = .1, ** = .05, and *** = .01}\label{tab:model}

% and this in place of [ht]: [!htb]
% ----------------------------------------------------------------------
\section{Discussion}\label{sec:disc}

This study makes several key advancements from prior work. For one, this is the first analysis, to our knowledge, that comparatively evaluates the spatial equity of docked and dockless bikeshare systems in the United States. Rather than focusing on either docked or dockless systems alone, the juxtaposition of the two approaches allows us to model comparative inequities in spatial access to bikeshare. Further, our sample contains the greatest diversity of cities of existing research on bikeshare in the United States. With data from 74 cities throughout the United States on both docked and dockless systems, our sample represents a more complete population than that used in existing works. Lastly, our methods are both generalizable and adaptably implemented, allowing for similar analyses in both new geographic areas and on new dimensions of equity. With publicly and freely available data and source code, as well as a published and maintained \textit{R} software package, our work is not only reproducible but adaptable for future analyses.

The results of these analyses have several implications for development of bikeshare programs in the United States, though careful consideration is required to utilize these findings effectively. In one way, the lesson to draw from this comparison seems unclear, given that the docked model seems to offer greater spatial equity in regard to race, but the dockless model seems to be better for equality of spatial access along educational boundaries. However, the main effect for the percent non-white was not statistically significant (p = .06) in predicting the number of bikes in a census tract, so this dimension of equity might warrant less consideration than education and class, which are both statistically significant main effects in our models and are highly correlated with race in the United States. Given that the docked and dockless models do not seem to differ in spatial equity in regard to poverty, though,   educational inequities appear to be able to be most proactively addressed by the choice of docked or dockless models. Hence, we argue that the dockless model offers greater potential for encouraging spatial equity in access to bikeshare services than the docked model due to its more equitable distribution of bikes along educational boundaries.

Still, though, there are several limitations to our study. For one, our sample data is at the level of observation of bikes, rather than bikes over time, preventing analyses on the way bikes move throughout time, and whether patterns in their movements (with regard to equity) are dictated more by users themselves or rebalancing carried out by operators. In the same way, our sociodemographic data only captures the current state of communities, while one recent study accounted for gentrification risks \cite{mooney2019} and others have tracked bikes over time \cite{dechardon2016, wergin2017}. Another element of level of observation that limits our analysis is that of spatial aggregation; at the level of observation of census tracts, our data glosses over neighborhood-level (and smaller) differences in demography and geography, which have recently been shown to play significant roles in social access and mobility in general \cite{nyt2018}. On another note, our study only captures questions of spatial equity by education, race, and class---differential access to these services also arises from disparities in information, internet access, and other social factors \cite{howland2017, lee2017, mcneil2018, howland2018}. Research on the social dimensions of equity in bikeshare under the docked and dockless models is an important next step. Lastly, because the sample data comes from a large number of different operators, the explanations for why bikes are geographically distributed the way they are under the two models could differ slightly based on the policies and pricing plans of the systems. For instance, some programs with both docked and dockless options charge a fee for bikes not parked at docking stations, presumably concentrating free bikes in higher income areas, while some others incentivize rebalancing on the part of users by charging less for rides that end in higher-demand areas. On the whole, though, the sample captures statistically significant disparities in spatial equity between the two approaches.

Altogether, we find that dockless models operate more equitably than the traditional docked model along educational boundaries, but did not find significant disparities in access by class. Though programs operating under the docked model seem to operate more equitably with regard to race, our models did not capture aggregate inequities along these boundaries. A key area for future research centers on other dimensions of equity when comparing the docked and dockless models, as well as taking into account the movement of bikes over time.

\clearpage

\bibliographystyle{plain}
\bibliography{sources.bib}

\begin{thebibliography}{10}

\bibitem{nabsa_repo}
North American~Bikeshare Association.
\newblock {gbfs}: {A} {Standardized} {Data} {Feed} {for} {Bike} {Share}
  {System} {Availability}, 2015.

\bibitem{babagoli2019}
Masih~A Babagoli, Tanya~K Kaufman, Philip Noyes, and Perry~E Sheffield.
\newblock Exploring the health and spatial equity implications of the new york
  city bike share system.
\newblock {\em Journal of Transport \& Health}, 13:200--209, 2019.

\bibitem{nyt2018}
Emily Badger and Quoctrung Bui.
\newblock Detailed {Maps} {Show} {How} {Neighborhoods} {Shape} {Children} for
  {Life}.
\newblock {\em The New York Times}, 2018.

\bibitem{bhuyan2019}
Istiak~A Bhuyan, Celeste Chavis, Amirreza Nickkar, and Philip Barnes.
\newblock Gis-based {Equity} {Gap} {Analysis}: {Case} {Study} of {Baltimore}
  {Bike} {Share} {Program}.
\newblock {\em Urban Science}, 3(2):42, 2019.

\bibitem{chetty2018}
Raj Chetty, John~N Friedman, Nathaniel Hendren, Maggie~R Jones, and Sonya~R
  Porter.
\newblock The {Opportunity} {Atlas}: {Mapping} the {Childhood} {Roots} of
  {Social} {Mobility}.
\newblock Technical report, National Bureau of Economic Research, 2018.

\bibitem{conrow2018}
Lindsey Conrow, Alan~T Murray, and Heather~A Fischer.
\newblock An {Optimization} {Approach} for {Equitable} {Bicycle} {Share}
  {Station} {Siting}.
\newblock {\em Journal of transport geography}, 69:163--170, 2018.

\bibitem{dechardon2016}
Cyrille~M{\'e}dard de~Chardon, Geoffrey Caruso, and Isabelle Thomas.
\newblock Bike-share {Rebalancing} {Strategies}, {Patterns}, and {Purpose}.
\newblock {\em Journal of Transport Geography}, 55:22--39, 2016.

\bibitem{ellin2012}
Nan Ellin.
\newblock {\em Good {Urbanism}: {Six} {Steps} to {Creating} {Prosperous}
  {Places}}.
\newblock Island Press, 2012.

\bibitem{fishman2016}
Elliot Fishman.
\newblock Bikeshare: {A} {Review} of {Recent} {Literature}.
\newblock {\em Transport Reviews}, 36(1):92--113, 2016.

\bibitem{nabsa_gbfs}
Nicole Freedman and Mitch Vars.
\newblock {North} {American} {Bikeshare} {Systems} {Adopt} {Open} {Data}
  {Standard}.
\newblock 2015.

\bibitem{griffin2018}
Greg~P Griffin and Junfeng Jiao.
\newblock Crowdsourcing bike share station locations: Evaluating participation
  and placement.
\newblock {\em Journal of the American Planning Association}, pages 1--13,
  2018.

\bibitem{handy2005}
Susan Handy.
\newblock Smart {Growth} and the {Transportation}-{Land} {Use} {Connection}:
  {What} {Does} the {Research} {Tell} {Us?}
\newblock {\em International Regional Science Review}, 28(2):146--167, 2005.

\bibitem{hosford2018}
Kate Hosford and Meghan Winters.
\newblock Who are {Public} {Bicycle} {Share} {Programs} {Serving}? {An}
  {Evaluation} of the {Equity} of {Spatial} {Access} to {Bicycle} {Share}
  {Service} {Areas} in {Canadian} {Cities}.
\newblock {\em Transportation Research Record}, 2672(36):42--50, 2018.

\bibitem{howland2018}
Steven Howland, Nathan McNeil, Joseph Broach, John Macarthur, and Jennifer
  Dill.
\newblock Bike share and equity in low-income communities of color: What
  opportunities are there to include older adults?
\newblock Technical report, 2018.

\bibitem{howland2017}
Steven Howland, Nathan McNeil, Joseph~Paul Broach, Kenneth Rankins, John
  MacArthur, and Jennifer Dill.
\newblock Breaking {Barriers} to {Bike} {Share}: {Insights} on {Equity} from a
  {Survey} of {Bike} {Share} {System} {Owners} and {Operators}.
\newblock 2017.

\bibitem{opp_insights}
Opportunity Insights.
\newblock Table 9: {Neighborhood} {Characteristics} by {Census} {Tract}.
\newblock https://opportunityinsights.org/data/.

\bibitem{katz1994}
Peter Katz, Vincent Scully, and Todd~W Bressi.
\newblock {\em The {New} {Urbanism}: {Toward} an {Architecture} of
  {Community}}, volume~10.
\newblock McGraw-Hill New York, 1994.

\bibitem{kroman2017}
David Kroman.
\newblock Can {Seattle’s} {New} {Bike} {Share} {Succeed} where {Pronto}
  {Failed?}
\newblock {\em Crosscut}, 2017.

\bibitem{le2018}
Huyen~TK Le, Ralph Buehler, and Steve Hankey.
\newblock Correlates of the {Built} {Environment} and {Active} {Travel}:
  {Evidence} from 20 {U}{S} {Metropolitan} {Areas}.
\newblock {\em Environmental health perspectives}, 126(07):077011, 2018.

\bibitem{lee2017}
Richard~J Lee, Ipek~N Sener, and S~Nathan Jones.
\newblock Understanding the {Role} of {Equity} in {Active} {Transportation}
  {Planning} in the {United} {States}.
\newblock {\em Transport {Reviews}}, 37(2):211--226, 2017.

\bibitem{leister2018}
Emily~Hentz Leister, Nicole Vairo, Dangaia Sims, and Melissa Bopp.
\newblock Understanding {Bike} {Share} {Reach}, {Use}, {Access} and {Function}:
  {An} {Exploratory} {Study}.
\newblock {\em Sustainable cities and society}, 43:191--196, 2018.

\bibitem{wiki:bikeshare}
{List of {Bicycle}-{Sharing} {Systems}}.
\newblock List of {Bicycle}-{Sharing} {Systems} --- {W}ikipedia{,} the free
  encyclopedia, 2019.
\newblock [Online; accessed 9-May-2019].

\bibitem{litman2015}
Todd Litman.
\newblock {\em Evaluating {Active} {Transport} {Benefits} and {Costs}}.
\newblock Victoria Transport Policy Institute, 2015.

\bibitem{martin2014}
Elliot~W Martin and Susan~A Shaheen.
\newblock Evaluating {Public} {Transit} {Modal} {Shift} {Dynamics} in
  {Response} to {Bikesharing}: a {Tale} of {Two} {U}{S} {Cities}.
\newblock {\em Journal of Transport Geography}, 41:315--324, 2014.

\bibitem{mcneil2018}
Nathan McNeil, Joseph Broach, and Jennifer Dill.
\newblock Breaking {Barriers} to {Bike} {Share}: {Lessons} on {Bike} {Share}
  {Equity}.
\newblock {\em Institute of Transportation Engineers Journal}, 88(2):31--35,
  2018.

\bibitem{meng2018}
Chao Meng.
\newblock Evaluation of the {Equity} of {Bikeshare} {System} {Accessibility}: A
  {Case} {Study} of {Chicago}.
\newblock 2018.

\bibitem{midgley2011}
Peter Midgley.
\newblock Bicycle-{Sharing} {Schemes}: {Enhancing} {Sustainable} {Mobility} in
  {Urban} {Areas}.
\newblock {\em United Nations, Department of Economic and Social Affairs},
  8:1--12, 2011.

\bibitem{mooney2019}
Stephen~J Mooney, Kate Hosford, Bill Howe, An~Yan, Meghan Winters, Alon Bassok,
  and Jana~A Hirsch.
\newblock Freedom from the {Station}: {Spatial} {Equity} in {Access} to
  {Dockless} {Bike} {Share}.
\newblock {\em Journal of Transport Geography}, 74:91--96, 2019.

\bibitem{us2015}
US~Department of~Health, Human Services, et~al.
\newblock Step {It} {Up}! {The} {Surgeon} {General’s} {Call} to {Action} to
  {Promote} {Walking} and {Walkable} {Communities}.
\newblock {\em Washington, DC: US Dept of Health and Human Services, Office of
  the Surgeon General. http://www. surgeongeneral.
  gov/library/calls/walking-and-walkable-communities}, 2015.

\bibitem{pucher2011}
John Pucher, Jan Garrard, and Stephen Greaves.
\newblock Cycling {Down} {Under}: a {Comparative} {Analysis} of {Bicycling}
  {Trends} and {Policies} in {Sydney} and {Melbourne}.
\newblock {\em Journal of Transport Geography}, 19(2):332--345, 2011.

\bibitem{ricci2015}
Miriam Ricci.
\newblock Bike {Sharing}: {A} {Review} of {Evidence} on {Impacts} and
  {Processes} of {Implementation} and {Operation}.
\newblock {\em Research in Transportation Business \& Management}, 15:28--38,
  2015.

\bibitem{gbfs}
Kaelyn Rosenberg and Simon Couch.
\newblock gbfs: {Interface} with {General} {Bikeshare} {Feed} {Specification}
  {Files.} {R} {Package}, September 2018.

\bibitem{shaheen2014}
Susan~A Shaheen, Elliot~W Martin, Adam~P Cohen, Nelson~D Chan, and Mike
  Pogodzinski.
\newblock Public {Bikesharing} in {North} {America} {During} a {Period} of
  {Rapid} {Expansion}: {Understanding} {Business} {Models}, {Industry} {Trends}
  \& {User} {Impacts}, {M}{T}{I} report 12-29.
\newblock 2014.

\bibitem{sin2017}
Ben Sin.
\newblock China's {Innovative} {Smartbike} {Sharing} {Startups} are {Hitting}
  {Obstacles} at {Home} and {Abroad}.
\newblock {\em Forbes}, pages 07--05, 2017.

\bibitem{smith2015}
C~Scott Smith, Jun-Seok Oh, and Cheyenne Lei.
\newblock Exploring the {Equity} {Dimensions} of {U}{S} {Bicycle} {Sharing}
  {Systems}.
\newblock Technical report, Western Michigan University. Transportation
  Research Center for Livable Communities, 2015.

\bibitem{speck2013}
Jeff Speck.
\newblock {\em Walkable {City}: {How} {Downtown} {Can} {Save} {America}, {One}
  {Step} at a {Time}}.
\newblock macmillan, 2013.

\bibitem{ursaki2015}
Julia Ursaki, Lisa Aultman-Hall, et~al.
\newblock Quantifying the {Equity} of {Bikeshare} {Access} in {U}{S} {Cities}.
\newblock Technical report, University of Vermont. Transportation Research
  Center, 2015.

\bibitem{wang2018}
Jueyu Wang and Greg Lindsey.
\newblock Measuring {Equity} in {Bike} {Share} {Programs}: A {Case} {Study} of
  the {Twin} {Cities}.
\newblock Technical report, 2018.

\bibitem{wergin2017}
Jon Wergin and Ralph Buehler.
\newblock Where do {Bikeshare} {Bikes} {Actually} {Go}?: {Analysis} of
  {Capital} {Bikeshare} {Trips} with {G}{P}{S} {Data}.
\newblock {\em Transportation Research Record}, 2662(1):12--21, 2017.

\end{thebibliography}

\end{document}